\documentclass[referee]{aasstitre} 
\usepackage{epsfig}
\begin{document}
  \thesaurus{} 
   \title{Interpretation of the UV spectrum of some stars with little 
   reddening.}

   \author{Fr\'ed\'eric Zagury \inst{}  }

   \offprints{F. Zagury}

   \institute{Department of Astrophysics, Nagoya University,
              Nagoya, 464-01 Japan\\
              email: zagury@a.phys.nagoya-u.ac.jp
             }

   \date{December, 1999}
   

  \titlerunning{Extinction in the UV-II}
  
  \maketitle

 \begin{abstract}
The UV spectrum of a few reddened stars will be decomposed into two 
terms.
One is the direct starlight, $F_{\star,\lambda}^0e^{-\tau_\lambda}$, 
which is the product of the flux of the star corrected for 
interstellar extinction, $F_{\star,\lambda}^0$, and of the extinction $e^{-\tau_\lambda}$.
The second is starlight scattered by interstellar dust into the beam of the observation.
This excess of scattered starlight affects the FUV part of the 
spectrum ($\lambda<2200\,\rm\AA$).
The combination of both terms gives the shape of the UV spectrum of a 
reddened star, 
with its characteristic depression at $2200\, \rm \AA$.
 \end{abstract} 
   \section{Introduction}
This paper is the second of a serie dedicated to the interaction of 
starlight and interstellar grains in the UV and supported by the 
observations of the International Ultraviolet Explorer (IUE) satellite.

The properties of the interstellar grains can be studied either by their 
capacity to scatter starlight, by observations of reflection nebulae, 
or by their capacity to extinguish starlight, 
by observing a star through an interstellar cloud.
The former method allows the evaluation of the phase function and of the 
albedo of interstellar grains.
The latter gives more specific information on
the proportion of starlight to be absorbed or scattered at each wavelength.
If extinction is the only process involved, the curve which is 
obtained by dividing the
spectrum of a star by the spectrum of an unreddened star of 
same spectral type shows the relative capacity, from one wavelength to another,
of the grains to extinguish starlight in the direction of the star.
This is an intrinsic property of the interstellar grains present in the direction of the 
observation.

The most salient feature of the UV spectrum of a star is the 
$\rm 2200\,\AA$ bump which appears when there is interstellar matter between the star and 
the observer. 
From the standpoint of interstellar grain properties, and if no 
scattered starlight is introduced into the beam of the observation, 
this implies the existence of a particular class of grains, the bump 
carriers, which extinguish light at wavelengths close to $\lambda_b \sim 
2175\, \rm \AA$. 
Those particles have never been formally identified to date. Efforts to 
comprehend the variations of the stars UV spectrum in 
different interstellar environments, all of which have assumed starlight extinction 
as the only process involved, did not really succeed in bringing a 
global understanding of the UV extinction curve (see Savage et al. 
\cite{savage85}, Fitzpatrick~\& Massa \cite{fitzpatrick86} and 
\cite{fitzpatrick88}). 

In a preceding paper (Zagury \cite{zagury1}, paper~I hereafter) the 
existence of the bump carriers was questioned as these carriers do not 
affect the UV spectrum of reflection nebulae.
The UV spectrums of the bright nebulae presented in paper~I 
were all interpreted as the result of starlight scattered by 
interstellar grains with identical albedo and phase function
across the UV and in the different directions of space 
sampled by the nebulae. 
It was also noted that these properties of the grains may be identical in the 
optical spectral range.
No bump or other particular feature is created at $\rm 2200\,\AA$
in the spectrum of the light scattered by a nebula.

If the bump carriers are not present in nebulae, the presence of an additional 
component due to scattering becomes 
the most reasonable explanation of the bump (Bless~\& Savage 
\cite{bless72}, and the appendix of this paper).
To explain the variations of the surface brightness of 
a nebula in function of its distance $\theta$ from the illuminating star (paper~I), 
the interstellar grains must have a strong forward scattering phase function.
In paper~I it was found that the maximum surface brightness a nebula can reach 
varies as a power law $\theta^\alpha$ ($\alpha<-1$) of $\theta$.
A value $\alpha=-2$ was found.
Consequently, the starlight scattered into the beam of
observation at close angular distance to the star may be a 
non-negligible proportion of the direct starlight.
The scattered starlight will be more important in a wavelength range 
for which the scattering medium has an optical depth close to 
$\tau_{max}\sim 1-2$.

The present paper will further develop these ideas.
I will study the UV spectrum of a selected sample of stars with a 
bump and little reddening.
Contrary to previous studies which use the logarithm of the 
spectrums as a mean of obtaining the extinction curve, 
in this paper, the direct linear data will be used. 
This is necessary if the spectrums can be 
decomposed into two separate components, the direct starlight and the
scattered starlight.

The stars to be studied in this paper have been selected because of 
the simple and straightforward interpretation which can be given of 
their UV spectrum.
The specific aspect of the UV reduced spectrum of these stars, 
defined as the ratio of 
the star to an unreddened star of same spectral type spectrums (section~\ref{data}),  clearly 
separates two spectral regions.
The long wavelength part of the spectrum is correlated with the 
reddening of the star, $E(B-V)$, and, for some stars, it is 
fitted down to the bump spectral region by an exponential 
of $1/\lambda$.
The exponential will be interpreted as the extinction of starlight 
$e^{-\tau_\lambda}$, where $\tau_\lambda$ is a linear function of 
$1/\lambda$ (section~\ref{expdec}).
The linear in $1/\lambda$ dependence of $\tau_\lambda$ was 
established in the optical in the 1930's (Greenstein \& Henyey 
\cite{greenstein41} and references therein) and detailed in more recent 
studies by Rieke~\& Lebofsky (\cite{rieke85}) and Cardelli, Clayton 
and Mathis (\cite{cardelli89}).
For the stars I have selected, this law extends to the UV.

At shorter wavelengths, $\lambda<\lambda_b$, 
a bump-like feature appears, superimposed to the exponential decrease.
This feature will be analysed in section~\ref{fuv} and attributed to 
additional starlight scattered by interstellar dust at very small 
(compared to the beam of the observations) angle to the star.
Hence, the $\rm 2200\,\AA$ bump is no longer considered as a depression, 
but rather the point at which scattering becomes noticeable.

The consequences of this interpretation are discussed in the conclusion.
     \section{Data} \label{data}
The IUE experiment (Boggess et al. \cite{boggess78a}, 
\cite{boggess78b}) and the process employed to obtain the final
IUE spectrums have been described in paper~I.
     
This paper will be concerned with seven stars, listed in 
table~\ref{tbl:star}.
Two of these stars, HD23480 (Merope) and HD200775 illuminate the Merope 
nebulae and NGC7023, which have been studied in paper~I.
For each star, the 
spectrum presented in the paper is an average of the best observations of the object. 
Some stars having been observed many times, only a few observations were 
necessary to ensure sufficient accuracy.
At times I used high dispersion spectrums and decreased the 
resolution by a median filter. 

\begin{figure*}[p]
\resizebox{!}{\textheight}{\includegraphics{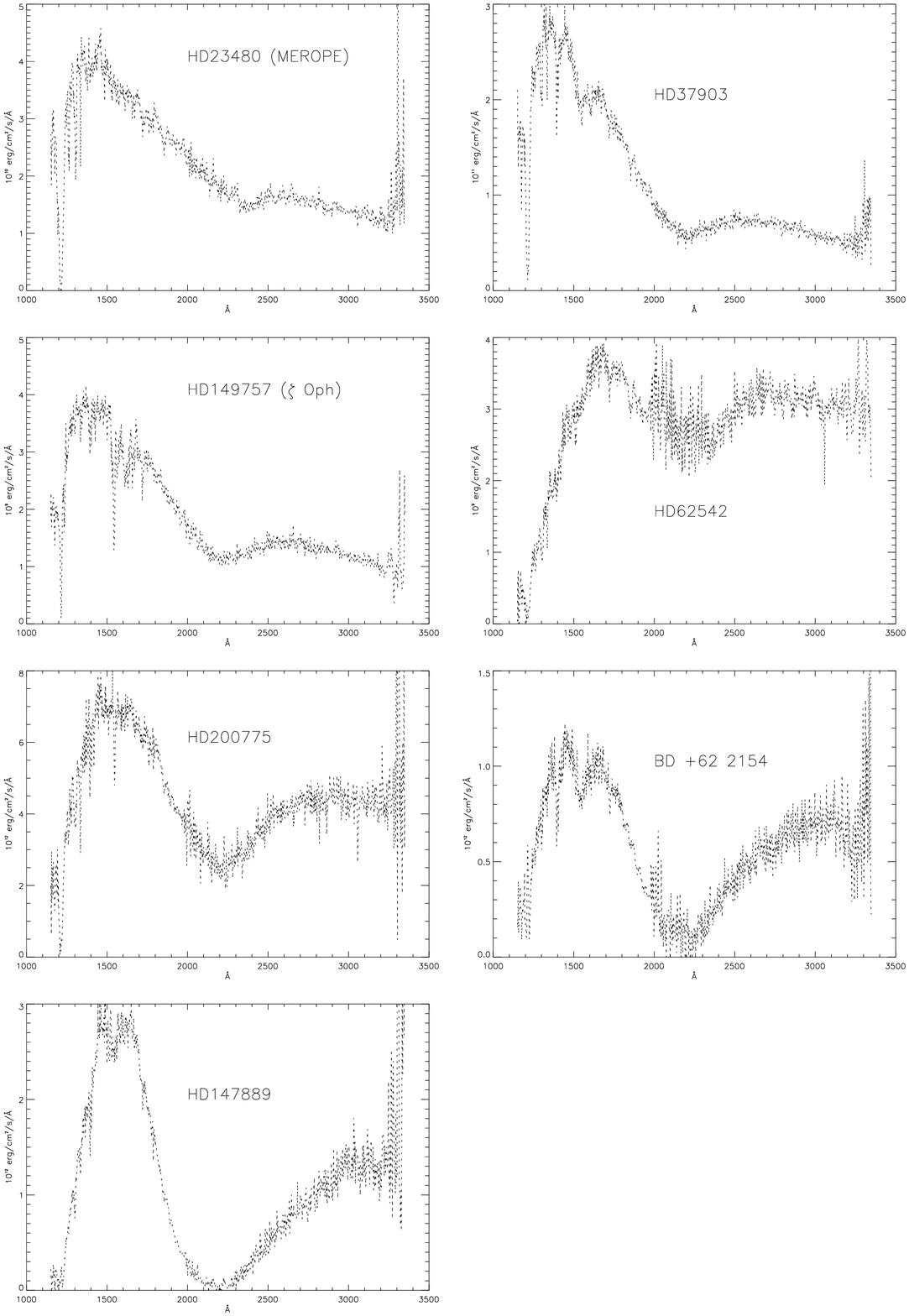}} 
\caption{UV spectrum of the reddened stars used in the paper.}
\label{etpresfig}
\end{figure*}

The spectrum of the seven stars selected for this study are 
presented in figure~\ref{etpresfig}, as they can be seen at IUE website.

The ratio of a reddened star spectrum to the spectrum 
of an unreddened star of same spectral type (reference star) will be 
called
a `reduced spectrum' of the reddened star. 
Each star has many reduced spectrums.
All reduced spectrums of a star are proportional.

The reduced spectrum of the seven stars, scaled to a common value at 
$3\,\mu\rm m^{-1}$, are presented in figure~\ref{etfig}.

The `absolute reduced spectrum' of a star is the reduced spectrum of 
the star obtained if the reference star is the star itself, corrected for reddening.  
It characterizes the interstellar medium between the star and the 
observer (section~\ref{diffspec}).
The logarithm of the absolute spectrum of a star is proportional to the 
extinction curve, $A_\lambda$ versus $1/\lambda$, in the direction of 
the star.

The reference stars which have been used to establish the 
reduced spectrum of the stars are listed in table~\ref{tbl:refstar}.  

A star spectrum can be presented as a 
function of the wavelength $\lambda$  or as a function of $\rm 
\lambda^{-1}$. 
The former manner keeps close to the data, whereas the 
latter is preferable since the optical depth varies as $1/\lambda$.
The latter presentation, more useful when dealing with 
scattering, has been chosen for most plots.
     \section{The $2200\,\rm \AA$ bump and the extinction of starlight} \label{diffspec}
When only extinction affects the transport of light between a star 
and the observer the spectrum of a star is the product of two 
terms: the unreddened flux of the star $F_{\star,\lambda}^0$ and the 
extinction $e^{-\tau_\lambda}$, where $\tau_\lambda=0.92A_\lambda$ is the optical depth at wavelength $\lambda$ of the medium 
in front of the star.
A reduced spectrum of the star, 
proportional to $e^{-\tau_\lambda}$, depends only on the interstellar 
medium between the star and the observer.
The absolute reduced spectrum of the star is $e^{-\tau_\lambda}$.

The presence of the bump has complicated the 
interpretation of the UV spectrum of the stars but has not changed
the ideas presented so far:
the bump is usually considered as a particular feature in the $\tau_\lambda$ 
function.

\begin{figure}
\resizebox{\textwidth}{!}{\includegraphics{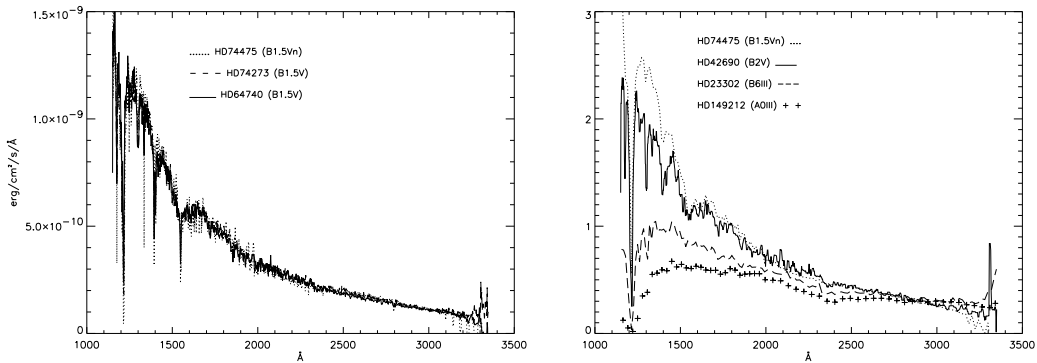}} 
\caption{\emph{Left}: Spectrums of unreddened stars with close spectral type. 
\emph{Right}: Variations of the UV spectrum of unreddened stars according to 
the spectral type. } 
\label{standart}
\end{figure}

The bump was proven to originate in the interstellar medium and 
related to the quantity of interstellar matter in front of the star
(Savage \cite{savage75}, see also Savage, Massa and Meade \cite{savage85}).
Stars of same spectral type with a bump have significant differences in their U.V. spectrums 
while stars with no bump and close spectral types superimpose very well 
after multiplication by an appropriate factor (figure~\ref{standart}). 
\section{The exponential decrease} \label{expdec}
\begin{figure*}
\resizebox{\hsize}{!}{\includegraphics{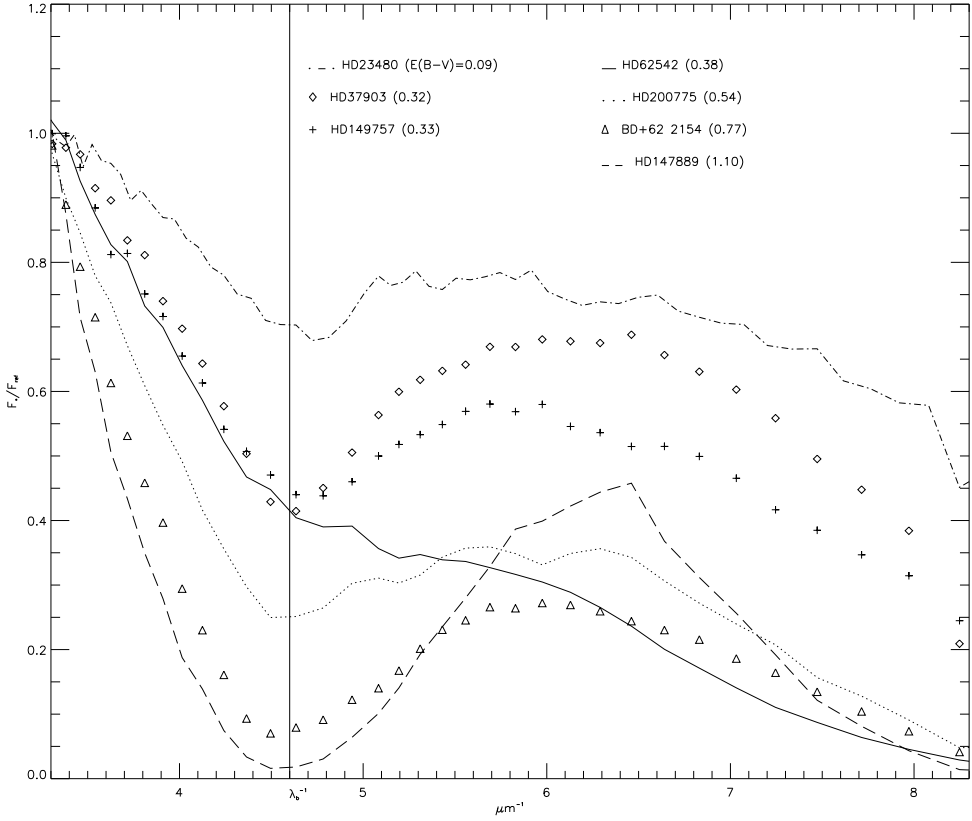}} 
\caption{Reduced spectrums of the stars divided by their value 
at $\lambda^{-1}=3\,\rm\mu m^{-1}$.}
\label{etfig}
\end{figure*}
\begin{figure*}
\resizebox{\hsize}{!}{\includegraphics{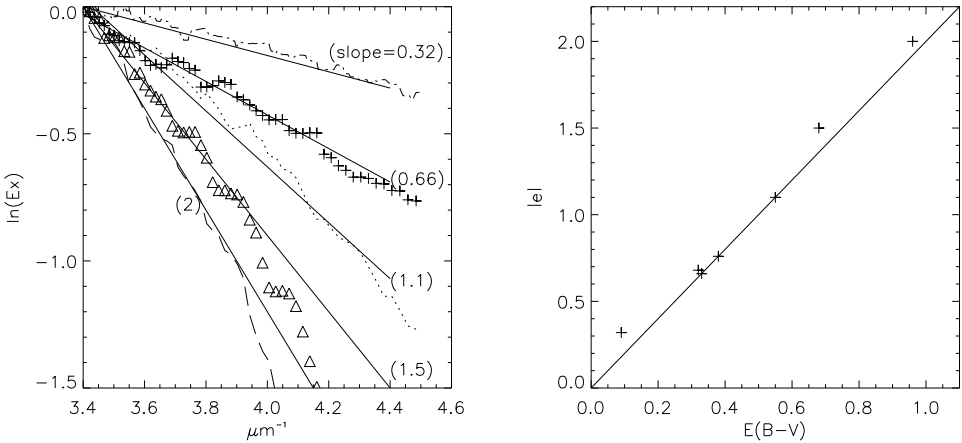}} 
\caption{\emph{left plot}: Logarithm of the reduced spectrum of the
stars of figure~\ref{etfig} in the long wavelength range.
\emph{right plot}: Correlation between $E(B-V)$ and the exponent of the 
exponential decrease. The solid line has a slope of $2$.}
\label{expfig}
\end{figure*}
Figure~\ref{etfig} plots the reduced spectrums of the seven selected stars. 
The reduced spectrums are smoothed by a median filter and 
scaled to have the same value at $3\,\rm\mu m^{-1}$. 
Stars are listed on the figure by order of increasing $E(B-V)$, written after the star name.

Each spectrum consists of 2 parts, clearly separated at $\lambda_b$. 
The long wavelength part ($\rm 3\,\mu m <1/\lambda < 1/\lambda_b$) varies steeply with 
increasing $1/\lambda$.  
This decrease will be called the `exponential decrease'.

There is a close correlation between the $E(B-V)$ 
value of a star and the steepness of its' exponential decrease: stars with larger 
$E(B-V)$ have a more rapid decrease.

For all the stars of the sample, the exponential decrease is well 
fitted by an exponential, $Ex=\beta e^{-e/\lambda}$, of $1/\lambda$.
The exponent $e$ can be estimated by 
taking the logarithm of the reduced spectrums (figure~\ref{expfig}, 
left).
In some directions (HD147889, BD~$+62\,2154$), especially the 
directions of highest $E(B-V)$, the exponential fit is best at long 
wavelengths, close to the optical wavelengths. 
In other directions, e.g. the directions of lowest $E(B-V)$ (HD23480, 
HD37903, HD149757, HD62542, HD200775), the 
fit applies to the totality of the exponential decrease.

Figure~\ref{expfig}, right, plots $e$ as a function of $E(B-V)$.
Within the error margin of $E(B-V)$ (estimated to be $\sim 0.1$~mag)
and on the determination of $e$ ($\sim 5\% $), 
we have: $e \sim 2E(B-V)$. 
This relation is justified if the linear relation which holds 
between $A_\lambda=1.08\tau_\lambda$ and $1/\lambda$ in the visible 
(Cardelli, Clayton~\& Mathis
\cite{cardelli89}, Rieke~\& Lebofsky \cite{rieke85}) is extended to the UV.
In this case:
\begin{eqnarray}
 A_\lambda &\,=\,&
 \frac{E(B-V)}{\frac{1}{\lambda_B}-\frac{1}{\lambda_V}}
 (\frac{1}{\lambda}-\frac{1}{\lambda_V}) 
 +A_V \nonumber \\
 &\,=\,& 2.2E(B-V)(\frac{1\,\mu{\rm m}}{\lambda}+0.46(R_V-4))
  \label{eq:av} \\ 
 \tau_\lambda &\,=\,& 2E(B-V)(\frac{1\,\mu{\rm m}}{\lambda}
 +0.46(R_V-4))
 \label{eq:tau}
\end{eqnarray}
with $R_V^{-1}=(A_B-A_V)/A_V= \tau_B/\tau_V-1$.

In the spectral range where equation~\ref{eq:tau} applies, extinction 
decreases as $e^{-\tau_\lambda} \propto e^{-2E(B-V)/\lambda}$.
The $Ex$ functions which fit the near UV spectrum 
of the stars we are concerned with have an identical exponential 
dependence on $1/\lambda$ (exponent$\,=-2E(B-V)/\lambda$).
The $Ex$ function prolong the linear optical extinction in the UV.
Since there is no reason to suspect an abrupt change in the optical 
depth -more specifically in the constant term of equation~\ref{eq:tau}-
when going from the optical to the UV, relation~\ref{eq:tau} must hold, 
in the directions sampled by the stars, 
from the near infrared to the near UV ($\lambda>\lambda_b$).

The left plots of figures~\ref{spec1fig}, \ref{spec2fig} and 
\ref{spec3fig}-top represent a reduced 
spectrum in the directions where the exponential decrease can be 
fitted by the $Ex$ function down to $\lambda_b$. 
The solid line is the exponential fit.
The close fit the $Ex$ function provides to the exponential 
decrease indicates that, in these directions, the extinction curve
is a linear function of $1/\lambda$ 
according to equation~\ref{eq:av} from the optical to $\lambda_b$.
\section{Absolute reduced spectrums}\label{absspec}
\begin{figure*}[p]
\resizebox{!}{0.8\textheight}{\includegraphics{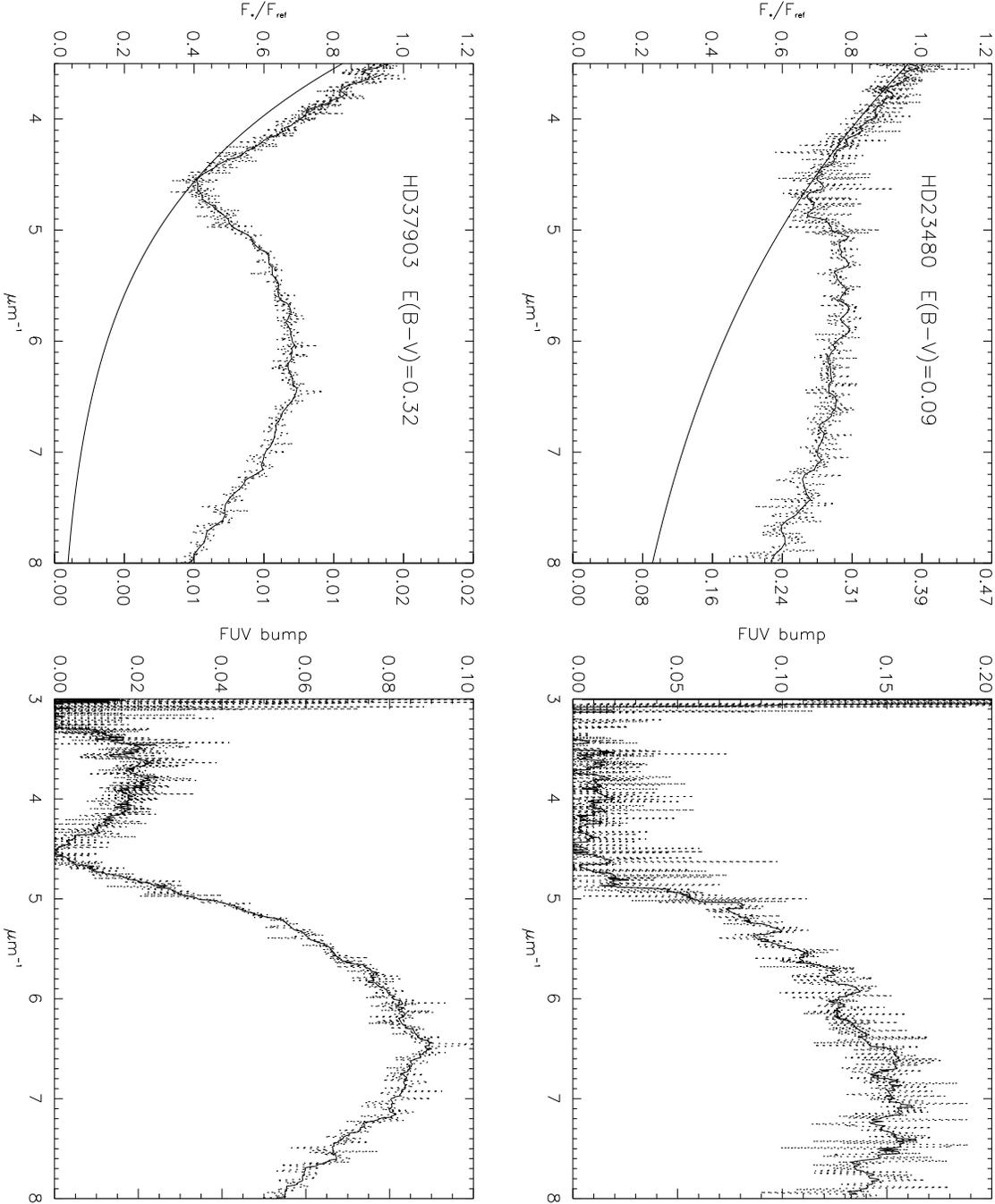}} 
\caption{Figure~\ref{spec1fig} to figure~\ref{spec3fig}.
\emph{left plot}: reduced spectrum of the star and best exponential 
fit to the long wavelength part of the spectrum. The curve in dots is 
the unsmoothed spectrum. The solid line spectrum was obtained by applying a 
median filter to the unsmoothed spectrum. 
The right y-axis is the absolute calibration of the 
plot, assuming $R_V=3$.
\emph{right plot}: Absolute reduced spectrum of the star minus the 
exponential decrease.}
\label{spec1fig}
\end{figure*}
\begin{figure*}[p]
\resizebox{!}{\textheight}{\includegraphics{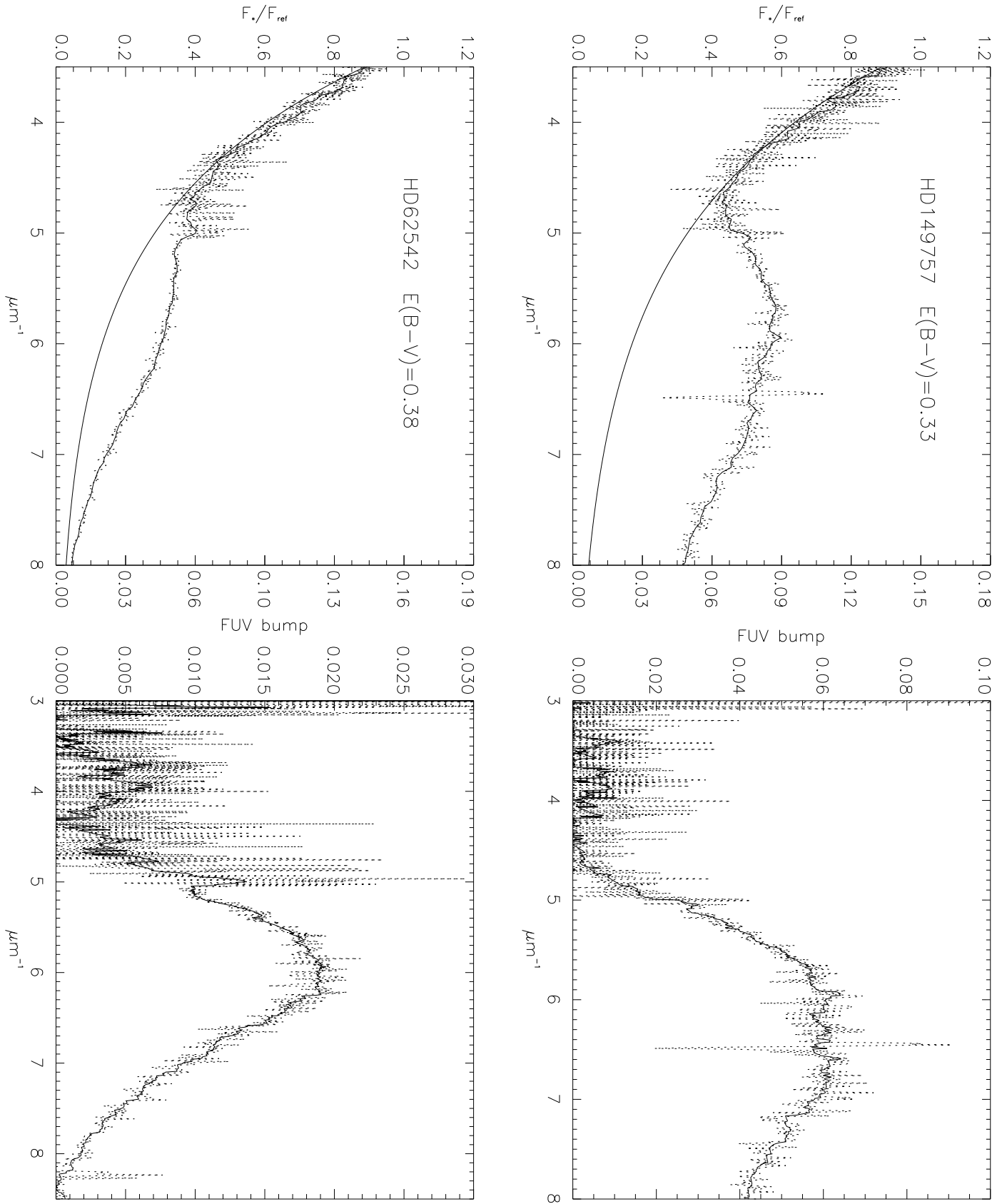}} 
\caption{Same as in figure~\ref{spec1fig}.}
\label{spec2fig}
\end{figure*}
\begin{figure*}[p]
\resizebox{!}{0.8\textheight}{\includegraphics{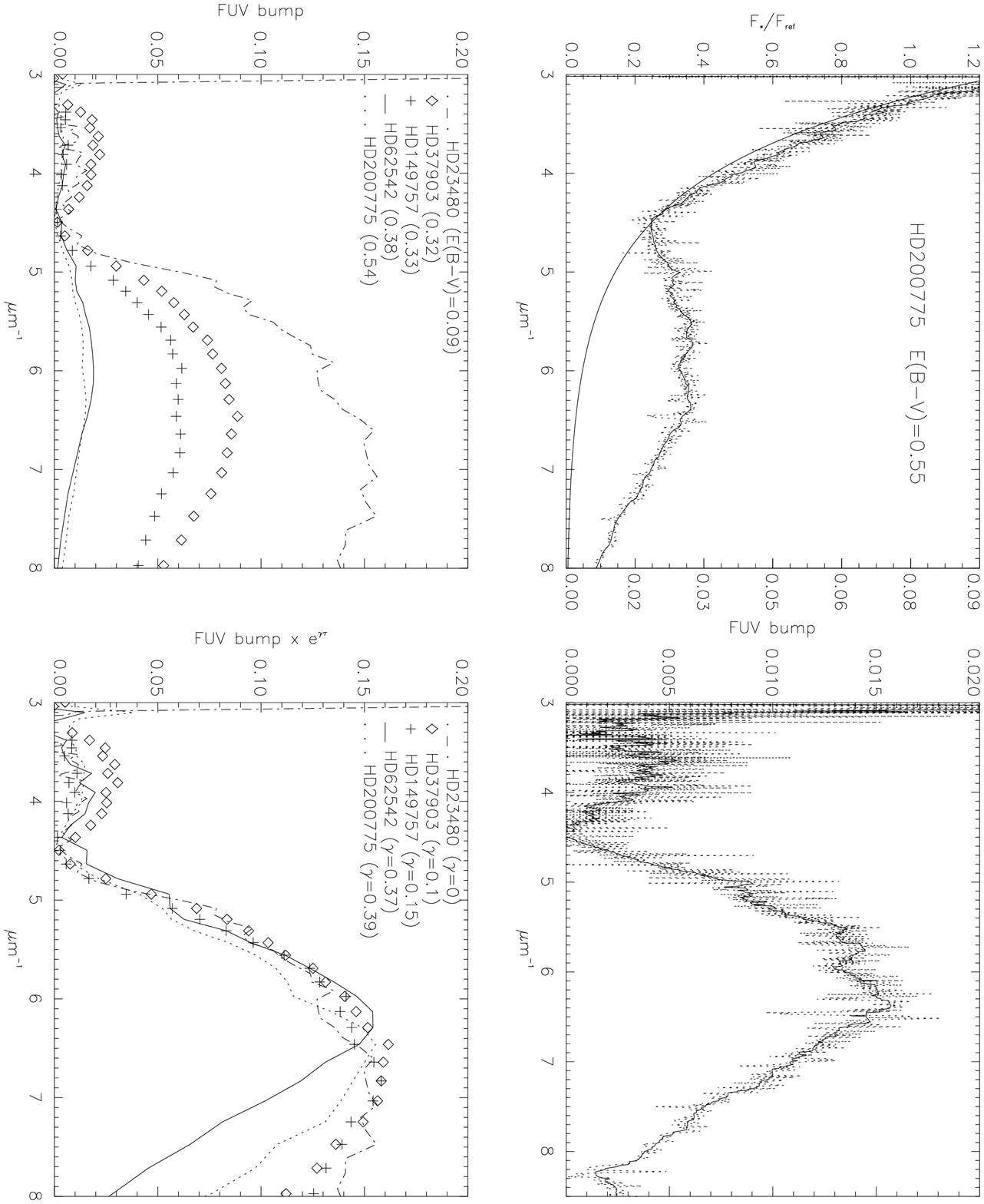}} 
\caption{
\emph{upper plots}: Same as in figure~\ref{spec1fig}.
\emph{bottom left}: The absolute reduce spectrum of the stars are 
presented on the same plot.
\emph{bottom right}: The absolute reduce spectrums are scaled to a 
similar maximum value ($\sim 0.15$). The spectrum are multiplied by 
an $e^{-\tau'_\lambda}$ function. $\tau'_\lambda$ was empirically 
determined using equation~\ref{eq:tau} and $R_V=3$.}
\label{spec3fig}
\end{figure*}
Knowledge of $E(B-V)$ along with relation~\ref{eq:tau} permits an 
exact scaling of the reduced 
spectrums in the directions where the exponential decrease is well 
fitted by an exponential function of $1/\lambda$.

In these directions, the exponential $Ex=\beta e^{-2E(B-V)/\lambda}$ multiplied by 
a scaling factor $\alpha_s$ must equal 
the extinction of starlight, $e^{-\tau_\lambda}$. 
$\alpha_s$ is unambiguously determined from equation~\ref{eq:tau} by:
\begin{eqnarray}
\alpha_s &\,=\, & \frac{1}{\beta}e^{-0.92E(B-V)[R_V-4]}\\
    &\,=\,&\frac{1}{\beta}e^{-0.92A_V(1-\frac{4}{R_V})}
    \label{eq:alphas}
\end{eqnarray}€
$\alpha_s$ is the factor to be applied to the star reduced spectrum in order 
to obtain the absolute reduced spectrum and the extinction curve in the 
direction of the star. 

In low density regions where $R_V$ is supposed to be $\sim 3$ 
(Cardelli et al. \cite{cardelli89}), we 
have:
\begin{eqnarray}
 \tau_\lambda &\,=\,& 2E(B-V)(\frac{1\,\mu{\rm m}}{\lambda}-0.46)
 \label{eq:taured} \\
\alpha_s &\,=\, & \frac{1}{\beta}e^{0.92E(B-V)}
    \label{eq:alphasred}
\end{eqnarray}€
With equation~\ref{eq:alphasred} the absolute reduced spectrum in the 
direction of a star can be deduced from a reduced spectrum and from the 
associated $Ex$ function.
This property was used to calibrate the right axis of the left hand 
plots and the right hand plots of 
figures~\ref{spec1fig} to \ref{spec3fig}.
\section{The FUV spectrum} \label{fuv}
\subsection{Analysis} \label{fuvan}
There is no evident correlation 
between $E(B-V)$ in the direction of a star and the short wavelength part 
($\rm 1/\lambda_b <1/\lambda < 8\,\mu m$) of its reduced
spectrum (figure~\ref{etfig}).  
The bump-like feature at $1/\lambda >1/\lambda_b$ is added 
to the tail of the exponential decrease. 
When the latter is slow, indicating small $E(B-V)$, the bump feature is tilted 
because of the underneath exponential decrease. 
HD23480, HD149757, HD62542, HD200775 are typical examples (see 
figure~\ref{etfig} and the left plots of figures~\ref{spec1fig} and \ref{spec2fig}).

In the directions of the stars selected in figure~\ref{spec1fig} to figure~\ref{spec3fig}, 
the $Ex$ function fits the exponential 
decrease down to $\lambda_b$, showing no excess of extinction or other
peculiarity at this wavelength.
In all directions the reduced spectrum seems to catch up with the 
exponential decrease for large $\tau_\lambda$-values.

The reduced spectrum of the stars, right hand plots of figure~\ref{spec1fig} 
to figure~\ref{spec3fig}, comprises the expected exponential 
extinction of starlight and the additional bump-like feature at short 
wavelengths.
The bump feature can be isolated if, for each direction,
$e^{-\tau_\lambda}$ is substracted from
the absolute reduced spectrum in the same direction.
The resulting curves are plotted at the right hand of figure~\ref{spec1fig}  to 
\ref{spec3fig}.

The short wavelength bump in the different directions are 
represented on the same plot, figure~\ref{spec3fig}, down and left.
With increasing $E(B-V)$, the height of the bumps tends to decrease 
and the short wavelength decrease (high $\tau_\lambda$) is steeper.
\subsection{Scattering}\label{fuvsca}
Scattering was ruled out as an explanation of the $2200\,\rm \AA$ bump 
for reasons which are summarized in the appendix and are questionable. 
The spectrum of nebulae (paper~I) did not reveal 
the presence of the bump carriers in the interstellar medium.
The exponential decrease of the UV spectrums can also logically be 
interpreted as the extinction of starlight by the same mean grain 
population responsible of the `normal' extinction process of starlight. 
For the stars studied here there does not seem to be any peculiarity 
at $\lambda_b$, and there is no need of a particular class of grains which 
would extinguish starlight at this wavelength.
The bump-like FUV feature is superimposed on the exponential decrease 
and appears as an additive feature to the extinction of the direct 
starlight $e^{-\tau_\lambda}$. 
It must arise (Bless et Savage \cite{bless72} and the 
Appendix, this paper) as a result of the introduction of scattered 
starlight into the beam of observation. 

If scattering is added to direct starlight, the light we receive from 
the direction of a star is the sum of the direct starlight, $F^0_{\star,\lambda} 
e^{-\tau_\lambda}$, and of the scattering component $F^0_{\star,\lambda} 
Sca_\lambda 
e^{-\tau'_\lambda}$. $F^0_{\star,\lambda}$ is the stellar flux at 
$\lambda$ corrected 
for reddening, $Sca_\lambda$ is the proportion (relative $F^0_{\star,\lambda}$) 
of starlight at wavelength $\lambda$ scattered into the beam.
$\tau'_\lambda<\tau_\lambda$ is the optical depth which accounts for the extinction of 
starlight between the star and the scattering medium and for the 
extinction of the scattered light.

The $Sca_\lambda$ function depends on the structure of the medium 
sampled by the line of sight.
The light scattered by a medium made of clumps with low density will not 
have the same spectrum as the light scattered by high density clumps.
The small scale structure of the medium, that is the density 
distribution of the different regions which compose the scattering 
medium at scales probably smaller than the IUE beam (Falgarone et al 
\cite{falgarone}), determines the shape of the 
$Sca_\lambda$ function.

With increasing $E(B-V)$ the FUV bump has a smaller maximum and 
a steeper FUV decrease (left-bottom plot of figure~\ref{spec3fig}).
This can be attributed 
to the $e^{-\tau'_\lambda}$ term of the scattered emission which 
affects the shortest wavelengths.

In the right hand bottom plot of figure~\ref{spec3fig} the FUV bumps 
are multiplied by an appropriate 
function $e^{-\tau'_\lambda}=e^{-\gamma \tau_\lambda^0}$, $\tau_\lambda^0$ given by 
equation~\ref{eq:taured} with $E(B-V)=1$, and $\gamma$ an appropriate 
constant. 
$\gamma$ is adjusted for all the curves to have comparable maximums.
All the curves of the figure have an identical growth between 
$\lambda_b$ and the maximum.

The curves for the direction of lowest reddening (HD23480, HD37903, 
HD149757)  superimpose well, 
supporting the idea of scattering by a medium with similar 
characteristics in these directions:
the $Sca_\lambda$ functions are identical and the scattered component of the 
absolute reduced spectrums differ by the extinction term $e^{-\tau_\lambda'}$ only.

The curves for the directions of HD200775 and HD62542 
cannot be superimposed to the others, thus differing by the 
$Sca_\lambda$ function, which indicates an environment of different nature.
\subsection{Implications}\label{scacond}
If scattering is responsible for the FUV bump of a star, 
the bottom plots of figure~\ref{spec3fig} show 
that it can represent up to $\sim 15 \%$  
of the direct starlight received on earth and corrected for 
extinction.

Most of the additional light has to be scattered within 
the $\theta_0=1"$ angle of the small $S$ aperture of the IUE telescope since 
the differences between observations made with the $S$ and 
with the large $L$ apertures of the IUE telescop are explained by 
pointing problems (section~\ref{asca}).
Because the ratio of $L$ to $S$ observations is generally greater 
than $1.2$, the difference of the amount of scattered starlight between the two types of observations 
must be less than $20 \% $.

Thus, if scattering is responsible for the FUV bump of a star, 
the scattered light is necessarily an appreciable 
proportion of direct starlight and scattering by 
interstellar dust must be strongly oriented in the 
forward direction.

\subsection{Case of a $\theta^{-2}$ dependence of 
the maximum surface brightness of nebulae} \label{scacond1}
Suppose the $\theta^{-2}$ 
dependence of the maximum surface brightness of a nebula (paper~I) on 
angular distance $\theta$ to the illuminating star is verified 
and applies to very small angles, down to a lower limit $\theta_{min}$.
According to the equation~6 of paper~I, the amplitude of the FUV bump, 
$15 \% $ of the 
unreddened flux of the star measured on earth, will be justified if:
\begin{eqnarray}
    \theta_{min} &< &\theta_0 e^{-\frac{0.07}{\pi c}}  \label{eq:cond1}\\
\theta_{min} &< & 6\,10^{-4}\, "
\nonumber
\end{eqnarray}
The latter inequality assumes $c\sim 3\,10^{-3}$ and $\theta_0=1\,"$.
Note that this upper estimate of $\theta_{min}$ is extremely sensitive 
to the $c$ parameter.
A value of $c=15\,10^{-3}$, within the possible range of values 
observed in paper~I, gives $\theta_{min}<0.2\,"$. $c=10^{-3}$ 
gives $\theta_{min}<2\,10^{-10}\,"$.

If $r_a$ is the ratio of the maximum light received from the direction of a star 
observed with apertures $\theta_1$ and $\theta_0$:
\begin{eqnarray}
r_a &=& \frac{\pi c (1+2\ln (\theta_1/\theta_{min}))}
{\pi c (1+2\ln (\theta_0/\theta_{min}))}
\nonumber \\
&\sim &\frac{\ln (\theta_1/\theta_{min})}
{\ln (\theta_0/\theta_{min})}
\nonumber \\
&\sim& 1- \frac{\ln \theta_1}{\ln \theta_{min}}
\label{eq:cond2}
\end{eqnarray}
with $\theta_1$ and $\theta_{min}$ in arcsecond.

If $\theta_1 \sim \, 10"$, corresponding to the large aperture of the IUE telescope, and 
$\theta_{min}$ is of order $10^{-\alpha}\,"$, equation~\ref{eq:cond2} 
has the simple form:
\begin{equation}
    r_a=1+\frac{1}{\alpha}
    \label{eq:cond2red}
\end{equation}
Within this framework, conditions~\ref{eq:cond1} and 
\ref{eq:cond2} refine the validity domain of the $\theta^{-2}$ law.
Condition~\ref{eq:cond1} must be satisfied to account for the amount 
of scattered light which is observed.
The power received in the direction of a star will not differ from the 
small ($\theta_0 \sim 1\,"$) to the large ($\theta_1 \sim 10\,"$) 
aperture of the IUE telescope if $\alpha\gg 1$ (equation~\ref{eq:cond2red}).
$r_a$ less than $1.2$ will be achieved for $\alpha \ge 5$, i.e. 
$\theta_{min}\le 10^{-5}\,"$.

In general, for observations made with an aperture $\theta_1\sim 
10^{\alpha_m}\,"$, and provided that the $\theta^{-2}$ dependence of the surface brightness 
 extends to $10^{\alpha_m}\,"$, $r_a$ will be 
$1+\alpha_m/\alpha$.
This result also depends upon the filling factor of the scattering medium.

\section{Conclusion}\label{con}
The UV spectrum of selected stars with a bump and moderate reddening was 
decomposed into two parts. 
The first part is the expected extinction of 
starlight, $e^{-\tau_\lambda}$, with $\tau_\lambda$ a linear function of $1/\lambda$.
The second is the FUV bump, observed in figure~\ref{etfig}, which is 
superimposed on the tail of the exponential decrease.
This additional component was interpreted as starlight scattered by interstellar 
dust at very small angle to the star.

This decomposition is justified by the very close  
exponential fit which can be applied to the long wavelength exponential decrease of the 
spectrums (figure~\ref{spec1fig} to \ref{spec3fig}).
The exponent of the exponential is $e=2E(B-V)/\lambda$, as in the optical.
The exponential fit extends to the UV the linear 
relation between $A_\lambda$ and $1/\lambda$ which applies in the optical. 
There is no excess of extinction at $2200\,\rm \AA$ in the spectrum of the 
stars which have been selected.

The additive nature of the FUV bump follows from the analysis of 
figure~\ref{etfig} carried out in sections~\ref{expdec} and \ref{fuv}.
Its' interpretation  as scattered 
starlight is then the most credible one.
This interpretation provides a simple explanation of the similarities 
and of the differences between the bumps in the different directions 
(section~\ref{fuvsca}).
If scattering is present in the spectrum of the stars, most 
of it must occur at less than $1\,"$ to the star. 
It implies a strong forward scattering phase function of the 
interstellar grains, as found in paper~I.
An attempt to justify the amount of scattered light ($\sim 15\% $ of 
the star flux measured on earth and corrected for reddening) was 
proposed in section~\ref{scacond1}. 
It involves the power law found in paper~I, which need to be 
confirmed, of the variation of a nebula maximum surface brightness with 
angular distance to the illuminating star.

This interpretation of the UV spectrum of the stars has two important consequences.

If the UV spectrum of some stars with a bump is explained without requiring an 
excess of extinction from the bump carriers at $\lambda_b$, all the 
stars' UV spectrum must have a similar interpretation.
There is no reason why scattering should affect the spectrum of some 
stars solely and/or why the bump carriers should be present 
in some nebulae only.

In the directions of the selected stars, 
the extinction curve, $A_\lambda$ as a function of $1/\lambda$, is a straight line 
from the near infrared to the FUV.
The value of $A_\lambda$ at wavelength $\lambda$ is given by the 
equation~\ref{eq:av}.
If, as it was suggested in paper~I, grains have the same 
properties in all directions of space, this law must hold for all 
directions.

Why is scattering so important in the FUV?
A first reason comes from the extinction of starlight which is 
increased when moving to the shortest wavelengths.
Consequently, at short wavelengths, the scattered starlight will be a larger part of the 
total light received in the direction of a star.
It is also plausible that the structure of the interstellar medium 
favors scattering at particular wavelengths.
A medium constituted of small clumps of similar $A_V$ will scatter 
starlight in a particular wavelength range: high $A_V$ clumps will 
preferentially scatter toward the red while very low column density 
clumps will scatter in the UV.
Although this aspect of scattering was not developped here, it may 
be an important step in the comprehension of the spectrum of 
the stars.
\clearpage
\clearpage
\begin{table*}[tp]
       \[
\begin{tabular}{|l|l|l|c|c|c|c|c|c|c|c|}
\hline
name& $\alpha_{1950}$ &$\delta_{1950}$&l&b&$B$&$B-V$&$E(B-V)$&sp. 
type &$Par^{\,a1}$&$\star_{st}^{\,a2}$  \\
\hline
BD$\,+62\,2154$&$22\,58\,33$&$+63\,14\,52$&110.94&+03.26&9.76&0.43&0.68&B1V&&6 \\
HD147889&$16\,22\,23$&$-24\,21\,07$&352.86&+17.04&8.66&0.71&0.96&B2III/IV&7.36&3 \\
HD149757 $\zeta$~Oph&$16\,34\,24$&$-10\,28\,02$&62.8&+23.59&2.595&0.017&0.33&O9V&7.12&1 \\
HD200775&$21\,01\,00$&$+67\,57\,55$&104.06&+14.19&7.73&0.31&0.55&B2Ve&2.33&5 \\
HD23480 
Merope&$03\,43\,21$&$+23\,47\,39$&$166.57$&-23.75&4.113&-0.051&0.09&B6IVe&9.08&2 \\
HD37903&$05\,39\,07$&$-02\,16\,58$&206.85&-16.54&7.91&0.07&0.32&B1.5V&2.12&6 \\
HD62542&$7\,40\,58$&$-42\,06\,37$&255.92&-09.24&8.21&0.18&0.38&B3V&4.06&4 \\
\hline
 \end{tabular}
    \]
\begin{list}{}{}
\item[$a1$] parallaxe in $marcsec$ measured by Hipparcos
\item[$a2$] associated standard star number. Refers to table~\ref{tbl:refstar}
\end{list}
\caption[]{ Stars used in the paper. Except for notified exceptions, all informations come from Simbad 
database (http://simbad.u-strasbg.fr). $(B-V)_0$ from FitzGerald 
\cite{fitzgerald70}}
		\label{tbl:star}
\end{table*}
\begin{table*}[]
       \[
\begin{tabular}{|l|l|l|c|c|c|c|c|c|c|c|}
\hline
N$^{\circ}$&name& $\alpha_{1950}$ &$\delta_{1950}$&l&b&$B$&$B-V$&$E(B-V)$&sp. 
type &$Par^a$  \\
\hline
1&HD214680&$22\,37\,00$&$+38\,47\,22$&96.65&-16.98&4.673&-0.2&0.1&O9V&3 \\
2&HD215573&$22\,45\,48$&$-80\,23\,19$&309.03&-35.53&5.190&-0.123&0.017&B6IV&7.35 \\
3&HD31726&$04\,55\,27$&$-14\,18\,28$&213.5&-31.51&5.928&-0.206&-0.034&B2V&3.28 \\
4&HD32630&$05\,03\,00$&$+41\,10\,08$&165.35&+00.27&3.012&-0.146&0.054&B3V&14.87 \\
5&HD58050&$07\,21\,37$&$+15\,36\,58$&202.53&+14.19&6.261&-0.187&0.053&B2Ve&0.67 \\
6&HD74273&$08\,39\,30$&$-48\,44\,36$&267.13&-04.27&5.69&-0.21&0.04&B1.5V&2.12 \\
\hline
 \end{tabular}
    \]
\begin{list}{}{}
\item[$a$] parallaxe in $marcsec$ measured by Hipparcos
\end{list}
\caption[]{ Standard stars used for the stars in table~\ref{tbl:star}. All informations come from Simbad 
database (http://simbad.u-strasbg.fr). $(B-V)_0$ from FitzGerald 
\cite{fitzgerald70}}
		\label{tbl:refstar}
\end{table*}

\appendix
     \section{The possible explanations of the $2200\,\rm \AA$ bump} 
     \label{bumpexp}
 Bless \& Savage 
(\cite{bless72}) review the possible causes of the bump: stars with 
a bump have a peculiar energy distribution; a large amount of starlight is 
scattered into the line of sight by interstellar dust; the 
extinction properties of the interstellar medium are particular in 
the UV. 

Because of its relation to $E(B-V)$, the 
bump does not originate in the stars' atmosphere, nor does it come 
from a special energy distribution. 

One explanation has been widely 
accepted and developped in all studies of the feature: the 
bump is due to a special class of interstellar grains which extinguish 
light at $1175\,\rm \AA$. 
According to studies of the bump in various 
environments (Savage \cite{savage75}, Jenniskens \& Greenberg \cite{jenniskens93}, Nandy et al. 
\cite{nandy76}), those grains are well 
 mixed (in all environments) to the large grains 
population  responsible for the `normal' extinction process of 
starlight. 

The last possibility, that scattered 
light enters into the beam, was ruled out for reasons to be discussed 
 in \ref{asca}. 
     \subsection{Arguments used against scattering} \label{asca}
3 types of arguments were used to rule out the 
possibility of scattering as an explanation for the bump.

According to Bless \& Savage (\cite{bless72}), Code has calculated 
that grains of albedo close to one are required to produce enough 
scattering to explain the UV extinction curve. However those 
calculations suppose isotropic scattering. The importance of forward 
scattering, and its' consequences for our interpretation of the UV 
spectrum of nebulae has been emphasized in paper~I. 
Introduction of forward scattering will 
invalidate Code's calculations.

Snow \& York (\cite{snow74}) have compared the spectrums of $\rm 
\sigma \, Sco$ from 2 different observations, each of which involves a 
camera of different aperture. The Wisconsin spectrometer aboard the 
Orbiting astronomical Observatory~2 (OAO-2) has a large aperture of 
$8'\times3^{\circ}$, while the Princeton OAO-3 has an entrance slit 
$0.3''\times 39''$, $10^3$ smaller than OAO-2. Thus, if the UV 
spectrum of stars with a bump was due to pollution by scattered 
nebular light, the authors expect to find differences between the 2 
spectrums.  
No such difference is observed.

Snow \& York's work imposed a serious constraint on interstellar grains but 
did not demonstrate the abscence of scattering in the UV spectrum of stars. If 
the phase function is strongly forward scattering, that is if the 
assymetry parameter $g_0$ is close to 1,  significant scattering will 
come only from directions very close to the star. Both observations 
used by Snow \& York have relatively large apertures, and will receive 
nearly equal amounts of scattered light.

Snow \& York's experiment can be repeated with IUE data. In most of 
the stars observed with both apertures of the IUE camera there is a 
scaling factor of $1.2$ to 
$2$ between the $L$ and $S$ aperture spectrums, regardless of the star reddening. 
This difference also affects 
unreddened stars and is probably due to the difficulty of holding the star 
within the beam when using the small aperture. Hence, if UV spectrums are affected by 
scattering, it must occur at angular distances less than $1.5"$ from the star.

Witt~\&~Lillie (\cite{witt73}) study the diffuse Galactic light (DGL)
spectrum from the OAO-2 satellite. 
The arguments of this paper 
rely on models of both the interstellar medium, assumed to be a 
plane parallel slab of uniform density, and of interstellar grains. 
The apparent disagreement between the model and the observations is 
attributed to changes of dust albedo with wavelength and to a pronounced 
minimun of the albedo at $2200\,\rm \AA$ ($\lambda_b$).
No DGL spectra is presented  in Witt~\&~Lillie's paper but their results 
 imply a pronounced minimum of the DGL at $\lambda_b$. 

IUE has observed over 400 off-positions, `IUE SKY', free of luminous objects. 
Many of the observations have some cirrus on their 
line of sight, the $100\,\mu$m IRAS emission can range from 1 to a few
$10\,$MJy/sr. The only region I have found with a reliable signal 
across the entire LWR camera wavelength 
range is spatially close to -and probably scatters the light of- the star cluster 
NGC1910. It has a level of $8\,10^{-14}$~$\rm erg/cm^2/s/ \AA$ and no bump.
In all other observations the emission shortward of 
$\rm 2600\,\AA$ has a very broad amplitude which can be attributed 
either to noise or to a very low level of signal.  
None of the spectrums, even when many of them are co-added, shows evidence 
for extinction at $\rm 2200\,\AA$.

{}

\end{document}